# A systems biology approach to COVID-19 progression in a population

Magdalena Djordjevic[1], Andjela Rodic[2], Igor Salom[1], Dusan Zigic[1], Ognjen Milicevic[3], Bojana Ilic[1], Marko Djordjevic,[2,*]

[1] Institute of Physics Belgrade, University of Belgrade, Serbia.
[2] Computational Systems Biology Group, Faculty of Biology, University of Belgrade, Serbia.
[3] Department for Medical Statistics and Informatics, Faculty of Medicine, University of Belgrade, Serbia.
[*] Correspondence: dmarko@bio.bg.ac.rs

**Abstract:**

A number of models in mathematical epidemiology have been developed to account for control measures such as vaccination or quarantine. However, COVID-19 has brought unprecedented social distancing measures, with a challenge on how to include these in a manner that can explain the data but avoid overfitting in parameter inference. We here develop a simple time-dependent model, where social distancing effects are introduced analogous to coarse-grained models of gene expression control in systems biology. We apply our approach to understand drastic differences in COVID-19 infection and fatality counts, observed between Hubei (Wuhan) and other Mainland China provinces. We find that these unintuitive data may be explained through an interplay of differences in transmissibility, effective protection, and detection efficiencies between Hubei and other provinces. More generally, our results demonstrate that regional differences may drastically shape infection outbursts. The obtained results demonstrate the applicability of our developed method to extract key infection parameters directly from publically available data so that it can be globally applied to outbreaks of COVID-19 in a number of countries. Overall, we show that applications of uncommon strategies, such as methods and approaches from molecular systems biology research to mathematical epidemiology, may significantly advance our understanding of COVID-19 and other infectious diseases.

## 1. Introduction

As the novel COVID-19 disease caused by the SARS-CoV-2 virus took the world by a storm, the new pandemic quickly gained priority in scientific research in a wide range of biological and medical science disciplines. Despite that their prior expertise was in unrelated research fields, many researchers have successfully adapted their approaches and methods to examine various aspects of this viral infection and, thus, contributed to finding the necessary solutions. The systems biology community is not an exception (Alon, Mino, & Yashiv, 2020; Bar-On, Flamholz, Phillips, & Milo, 2020; Magdalena Djordjevic, Djordjevic, Ilic, Stojku, & Salom, 2021; Eilersen & Sneppen, 2020; Karin et al., 2020; Saad-Roy et al., 2021; Vilar & Saiz, 2020; Wong et al., 2020): those involved in modeling the dynamics of biological systems at the molecular and cellular level can directly apply the similar methodology in epidemiological studying of the virus spread – and this exactly is the central point of the present paper. In particular, dynamic models of biochemical reaction networks, in which the reaction kinetics follow the law of mass action, are analogous to compartmental epidemiological models which, instead of concentrations of chemical species, track the prevalence of individuals in defined population classes over time (Voit, Martens, & Omholt, (2015)). Moreover, gene expression dynamics is usually a result of the interplay between the changing rate of cell growth, on which the global physiological rates of molecule synthesis and degradation depend, and complex transcription regulation (Marko Djordjevic, Rodic, & Graovac, 2019). Therefore, modeling dynamics of gene circuits implies combining kinetic models, often relying on the law of mass action, with appropriate nonlinear functions describing the regulation part. In the case of the COVID-19 epidemic, one can note that the virus transmission in a population, driven by the biological capacity of the particular virus in the given environment, is coupled with strong, time-dependent regulation, represented by the epidemic mitigation measures imposed by governments. These similarities between the modeled systems may facilitate the application of the systems biology techniques to the epidemiology field of research. In this paper, we will show how such an approach can be used to assess the basic parameters of the COVID-19 epidemic progression in a given population. In particular, we will use the analogy outlined above to study the COVID-19 spread in Mainland China and test the hypothesis about possible reasons for the uneven disease spread in China provinces.

Our interest in Mainland China infection progression comes from Figure 1. The progression seems highly intriguing, as Hubei (with only 4% of China population) shows an order of magnitude larger number of detected infection cases (Fig. 1A) and two orders of magnitude higher fatalities (Fig. 1B) compared to the *total* sum in all other Mainland China provinces. The epidemic was unfolding well before the Wuhan closure (with the reported symptom onset of the first patient on December 1st, 2019) and within the period of huge population movement, which started two weeks before January 25th (the Chinese Lunar New Year) (Chen, Yang, Yang, Wang, & Bärnighausen, 2020). As a rough baseline, a modeling study of the infection spread from Wuhan (Wu, Leung, & Leung, 2020) estimated more than $10^5$ new cases per day in Chongqing alone – instead, the actual (reported) peak number for *all* Mainland China provinces outside Hubei was just 831.

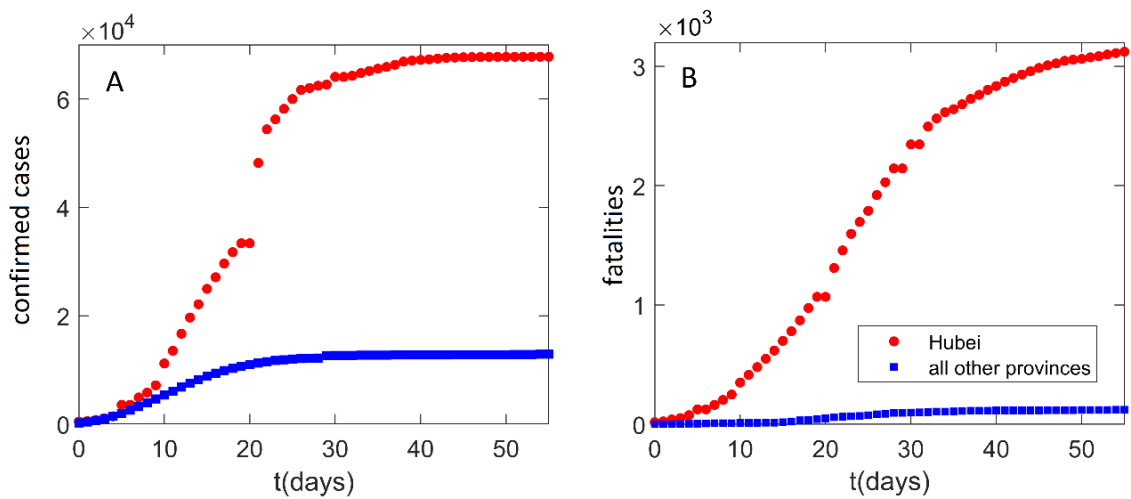

**Fig. 1. Infection and fatality counts for Hubei vs. all other provinces.** The number of (**A**) Detected infections, (**B**) Fatality cases. Zero on the horizontal axis corresponds to the time from which the data (Hu et al., 2020) are taken (January 23rd), which also coincides with the Wuhan closure. Red circles correspond to the observed Hubei counts. Blue squares correspond to the sum of the number of counts for all other provinces. The figure illustrates a puzzling difference in the number of counts between Hubei alone and the sum of all other Mainland China provinces.

Consequently, it is a notable challenge for computational modeling to understand drastic differences in COVID-19 infection and fatality counts observed between Hubei (Wuhan) and other Mainland China provinces. These drastic differences may be a consequence of an interplay between the virus transmissibility (influenced by environmental and demographic factors) and the effectiveness of the protection measures. Both can significantly change between different provinces (more generally different countries/regions), and the model has to infer this from available data (commonly the number of confirmed cases, publicly available for a large number of countries/regions).

The study presented here will therefore demonstrate the usefulness of the systems biology approach to the analysis of non-trivial COVID-19 data from China. In particular, the developed method will allow us to analyze the puzzling differences in dynamics trajectories in Mainland China provinces, and it will also turn out to be more generally applicable for understanding regional differences in outburst dynamics. The surprising differences in COVID-19 progression in different provinces may put strong constraints on the underlying infection progression parameters and allow us to understand:

i. What interplay between the inherent disease transmissibility and the effects of social distancing is responsible for the large difference in the count numbers between Hubei and the rest of Mainland China? Addressing this question in a proper way would make easier to comprehend how regional differences may shape the infection outbursts, which is important both locally (for explaining this puzzle), and more generally in the context of global COVID-19 pandemics progression.

ii. What is the Infected Fatality Rate (IFR, the number of fatalities per total number of *infected* cases) in China? Case Fatality Rate (CFR, the number of fatalities per *confirmed*/*detected* cases) can be obtained directly from the data but is highly sensitive to the testing coverage. IFR is a more fundamental mortality parameter, as it does not depend on the testing coverage, but is however much harder to determine, due to the unknown number of infected cases.

Addressing these questions allows understanding both the different response policies, and the inherent risks posed by the pandemics and will enable future cross-country comparisons. The developed methodology i) demonstrates the usefulness of applying transdisciplinary expertise to efficiently analyze problems of nationwide importance, ii) allows to readily analyze future outbreaks of COVID-19 and other infectious diseases, as it depends only on inference from straightforward and publically available data.

## 2. An overview of epidemics progression compartmental models

In epidemiology, for practical and ethical reasons, it is fairly impossible to conduct scientific experiments in controlled conditions in order to investigate the spread of the disease in the human population (Brauer, 2008). Therefore, epidemiologists usually resort to collecting data from clinical reports on the observed situation in the field and, then, using mathematical models to interpret these data, i.e., to infer the principles underlying the process of disease spreading. These principles may point to potentially successful control strategies, as well as to the probable future status of the disease in the population. Epidemiological data can often be incomplete or inaccurate due to poorly controlled or non-standardized collection methods, which significantly complicates modeling. However, even a qualitative agreement of the model with the data can provide useful information of great practical importance. Hence, model predictions are widely used for making various estimates and answering important questions about the seriousness of the epidemic consequences. For example, how many people will be infected, require treatment, or die, or how many patients should the public health facilities expect at any given time? Also, how long will the epidemic last? To what extent could quarantine and self-isolation of the infected contribute to mitigating the effects of the epidemic? Model predictions guide the development of strategies to control the epidemic spread, including vaccination programs.

When the goal is to discover the general principles of epidemic progression, simple mathematical models, which can be solved and analyzed with a "pencil on paper", are a logical choice as they give insight into the properties of the examined process despite failing to reproduce it in detail. In 1927, Kermak and McKendrick formulated a simple model that predicted behavior similar to that observed in numerous epidemics (Kermack & McKendrick, 1927). It was a type of compartmental model describing the infection spread in a population by analogy with a system of vessels connected by pipes through which a fluid flows. Namely, the population is divided into compartments, and assumptions are made about the nature and the rate of the flow between them. The structure of the compartmental model - which sections and how many of them it will contain and how they will be connected - depends on the characteristics of transmission of a given infectious disease and whether the past disease provides immunity to re-infections or not. The model set by these two scientists is known as SIR (from **S**usceptible - **I**nfected - **R**ecovered). It divides the population into three classes which correspond to compartments (Fig. 2): Susceptible (*S*) class includes healthy individuals susceptible to infection, which have never been exposed to the virus; Those who are infected and can infect others belong to the Infected (*I*) class; Removed (*R*) class encompasses those who are excluded from the population, either by quarantining the infected, or by acquiring immunity through recovery from disease or immunization, or by the death of the infected (Brauer, 2008).

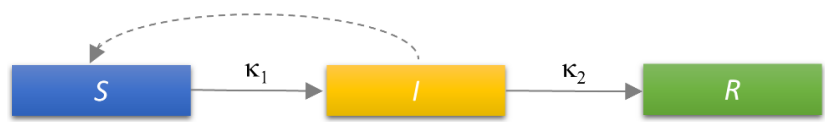

**Fig. 2. Schematic representation of the SIR model.** Rectangles denote model compartments containing susceptible (*S*), infected (*I*), and recovered (*R*) individuals in the population of size N. Permitted directions of flow between compartments are denoted by arrows, with the rates of flow indicated above them. The rates are expressed according to the law of mass action, where $\kappa_1$ and $\kappa_2$ are the rate constants. The dashed curve corresponds to bimolecular reaction, where newly infected are generated through interactions (contacts) between susceptible and already infected individuals.

Mathematically, this model is represented by a system of ordinary differential equations. The time derivative of the number of individuals in a compartment, i.e., the rate of their change, is given by the difference between the rates at which the compartment is filled and emptied. Analogous to the processes in which chemical species (e.g. proteins) are degraded or converted into others within a biochemical reaction network (Ingalls, 2013), the rate of transition of individuals from one compartment to another follows the law of mass action. For example, a person moves from compartment *S* to compartment *I* at the rate which is proportional to the product of the *S* and *I*, as the encounter with an infected person enables virus transmission to the susceptible one (Voit et al., (2015)).

By formulating such (or similar) models, one assumes that the epidemic is a deterministic process. Namely, the state of the population at all times is completely determined by its previous state and the rules described by the model. This is a reasonable approximation in cases where the numbers of individuals in the compartments are large, i.e., in a commonly considered deterministic range (>10). Such approximation (i.e., deterministic modeling) is well suited for the spread of COVID-19, which is up to now known for a large number of individuals in all compartments.

## 3. Systems biology approach to a compartmental model of the COVID-19 epidemic

The above-introduced SIR model is likely the simplest compartmental model in mathematical epidemiology and many subsequent models are derivatives of this basic form. Among others, these extensions have also been developed towards including control measures such as vaccination or quarantine (Diekmann, Heesterbeek, & Britton, 2012; Keeling & Rohani, 2011; Martcheva, 2015). However, COVID-19 brought a challenge to account for previously unprecedented social distancing measures, taken by most countries. When included, these effects have been, up to now, accounted for by the direct changes in the transmissibility term (Chowell, Sattenspiel, Bansal, & Viboud, 2016; Tian et al., 2020), which, however, corresponds to introducing a phenomenological dependence in otherwise mechanistic models. That is, to be included consistently in the model, social distancing should move individuals from one compartment to the other, just as vaccination and quarantine are usually implemented. On the other hand, it is necessary to construct a minimal mechanistic model in terms of the ability to explain the data with the smallest number of parameters, so that relevant infection progression properties can be inferred without overfitting. With this goal in mind, we used our systems biology background to develop a minimal model that accounts for all the main qualitative features of the SARS-CoV-2 infection spread under epidemic mitigation measures. As outlined above, we opt for a deterministic model due to the robust and computationally less demanding parameter inference (Wilkinson, 2018).

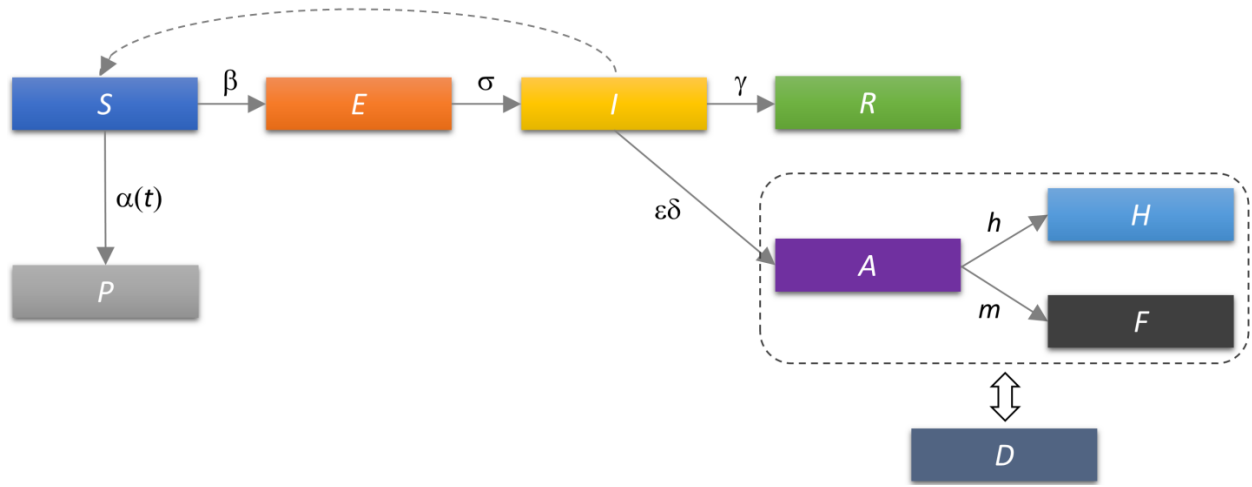

**Fig. 3. Schematic representation of the SPEIRD model.** Compartments and the transition rates are as indicated in the text, where transitions between different compartments are marked by arrows. The time-dependent transition rate from susceptible to protected category $\alpha(t)$ is indicated by the solid arrow. The infected can transition to the recovered category either without being diagnosed (transition to *R*), or being diagnosed and then transitioning to confirmed healed or fatality cases. The dashed rectangle indicates that *A*, *H*, and *F* categories in the starting model are substituted for the cumulative case counts (*D*), which removes *h* and *m* from the analysis, where *D* is fitted to the observed data.

To describe the COVID-19 epidemic, we developed SPEIRD model depicted schematically in Figure 3. It assumes that healthy persons susceptible to infection (*S*), can be infected, but in the case of this (and many other) viruses they do not immediately become contagious to other people, but first spend some time in the compartment *E* (**E**xposed to the virus) and then develop symptoms and pass to the compartment *I*. Infected persons can either recover at home, moving to the compartment R, or they can be diagnosed with SARS-CoV-2 virus infection (**A**ctive detected cases). *A* (Active) cases can, further, either become healed or die from the disease. To consistently implement the social distancing within this model structure, we included a compartment *P* (**P**rotected) in the model, which contains susceptible persons who are protected from exposure to the virus as a result of the epidemic mitigation measures, such as self-imposed isolation, social distancing, and advised changes in individual behavior.

The following differential equations describe how different categories change with time:

$$dS/dt = -\beta \cdot I \cdot S/N - \alpha(t) \cdot S \qquad (1)$$

$$dE/dt = \beta \cdot I \cdot S/N - \sigma \cdot E \qquad (2)$$

$$dI/dt = \sigma \cdot E - \gamma \cdot I - \varepsilon \cdot \delta \cdot I \qquad (3)$$

$$dA/dt = \varepsilon \cdot \delta \cdot I - h \cdot A - m \cdot A \qquad (4)$$

$$dH/dt = h \cdot A \qquad (5)$$

$$dF/dt = m \cdot A \qquad (6)$$

where $\beta$ is the infection rate in a fully susceptible population; $\alpha(t)$ - the time-dependent protection rate, i.e., the rate at which the population moves from susceptible to the protected category, quantifying the impact of the social protection measures. $\sigma$ – the inverse of the exposed period; $\gamma$ – the inverse of the infectious period; $\delta$ – the inverse of the period of the infection diagnosis; $\varepsilon$ - the detection efficiency; $h$ - the healing rate of diagnosed cases; $m$ - the mortality rate.

The probability that an infected person will meet a susceptible person is proportional to $S/N$, where $N$ is the total number of individuals in the population. The rate at which individuals move from $S$ to $E$ is obtained when the product of $I$ and $S/N$ is multiplied by the infection rate, $\beta$, which quantifies the efficiency of transmission of a particular virus in the population with certain demographic characteristics and meteorological conditions, and it does not depend on epidemic suppression measures. Thus, $\beta$ is a characteristic of the virus, the population, and the external conditions in which the virus is transmitted. Since the compartment $S$ is being emptied, the corresponding rate in the first equation is specified with the minus sign.

$S$ also decays by moving the individuals to $P$ with a protection rate that may vary with time. While mitigation measures are commonly accounted for by models with time-independent terms (Martcheva, 2015), we note that the social distancing term should depend on time, as this measure is introduced at a certain point in epidemics and may also evolve gradually. We denote the time point (more specifically, the date) of the onset of the social distancing measures in the examined population with $t_0$. The protection rate $\alpha(t)$ is then taken as 0 before $t_0$ and a constant value $\alpha$ afterwards.

One may notice a direct parallel between the model outlined above, and e.g., modeling gene expression regulation in systems biology with a step function that approximates the activity of a promoter to which repressor proteins are highly cooperatively bound: the promoter is initially silenced and upon receiving a signal which leads to the abrupt removal of repression, promoter activity rises sharply to its maximum value. We notice that the step function is a satisfactory approximation of the dynamics of social distancing, i.e., it may not be necessary to further increase the number of parameters by applying the Hill function (which describes a more gradual activation), since governments quickly introduced these measures, together with their effective implementation. Note however that in (Magdalena Djordjevic, Djordjevic, Ilic, Stojku, & Salom, 2021) we introduced a more complex model with Hill function, and provided analytical results for key properties of this model.

Compartment $E$ is filled by infecting the susceptibles and emptied by moving the individuals to $I$, with the rate σ representing the inverse value of the latent period during which the person is not contagious. While compartment $I$ is filled with individuals from $E$, it is depleted through two channels. Individuals move to $R$ with the rate γ, which is the inverse of the period of contagiousness, and to $A$ with the rate δ, which is the inverse of the time required for diagnosis, multiplied by ε, reflecting that only a fraction (likely small due to many asymptomatic infections) of the total infected are detected. Note that case detection reduces the number of individuals in $I$ that can infect susceptibles: the model assumes that the detected cases are quarantined and thus isolated from the general population. The numbers in compartments $A$, $H$, and $F$ change following the same logic described for the other compartments.

We can further simplify the analysis by looking at the total number of detected cases ($D$), which is the sum of $A$, $H$, and $F$. By adding the Eqs. (4)-(6), we obtain:

$$dD/dt = \varepsilon \cdot \delta \cdot I, \tag{7}$$

and thus lose two parameters, $h$ and $m$. The total number of detected cases in time is a measurable quantity from which we can determine the dynamics of other model compartments since this is the data that is available for various different regions and countries. Thereby, we assume that before $t_0$ social distancing does not take effect, and the measures introduced at $t_0$ will take effect on $D$~10 days later, as this is about the time that elapses between infection and detection/diagnosis (Feng et al., 2020). Consequently, for the first $t_0+10$ days, the $D$ curve reflects disease transmission without epidemic suppression measures.

### 3.1. *Virus transmission in the early stages of epidemics*

We will now focus on the dynamics of the infection spread at the very beginning of the epidemic, i.e., on the period before the introduction and practice of any control measures (Salom et al., 2021). Regarding the model, we assume that there is no social distancing (no transition from *S* to *P*), there is no quarantine, and almost the entire population consists of people susceptible to infection, so $S/N = 1$. This gives us an even simpler mathematical model which appears to be very useful because it allows analytical derivation of the expressions we need. Our system of Eqs. (1)-(3) and (7) is reduced to two linear differential equations that we can write in matrix form

$$\frac{d}{dt}\begin{pmatrix} E \\ I \end{pmatrix} = \begin{pmatrix} -\sigma & \beta \\ \sigma & -\gamma \end{pmatrix}\begin{pmatrix} E \\ I \end{pmatrix} = A\begin{pmatrix} E \\ I \end{pmatrix}, \qquad (8)$$

determine the eigenvalues and eigenvectors of the matrix and, subsequently, the solutions of the system, $E(t)$ and $I(t)$. Specifically, the cumulative number of infected in time, $I(t)$, is obtained according to the following equation:

$$I(t) = C_1 \exp(\lambda_+ t) + C_2 \exp(\lambda_- t), \qquad (9)$$

where λs are eigenvalues of the matrix. Since one of the eigenvalues, here denoted by $\lambda_-$, is negative, the corresponding term of the Eq. (9) will decrease over time, and $I(t)$ will be effectively described by the first term, already after few days from the epidemic outbreak (Salom et al., 2021). We can further derive this equation for the dependence of the logarithm of the number of detected cases in time:

$$\log(D(t)) = \log(\varepsilon \cdot \delta \cdot I(0)/\lambda_+) + \lambda_+ \cdot t. \qquad (10)$$

This is the straight line equation whose slope is given by the value of $\lambda_+$ (the dominant, positive eigenvalue of the matrix in Eq. (8)).

Once we know $\lambda_+$, we can calculate the value of the so-called basic reproduction number, $R_{0,\text{free}}$, by fixing mean values of the latency period and the infectivity period ($\gamma = 0.4$ days$^{-1}$, $\sigma = 0.2$ days$^{-1}$), which are known from the literature and characterize the fundamental infection processes (Kucharski et al., 2020; Li et al., 2020):

$$R_{0,\text{free}} = \frac{\beta}{\gamma} = 1 + \frac{\lambda_+(\gamma+\sigma) + \lambda_+^2}{\gamma\sigma}. \qquad (11)$$

$R_{0,\text{free}}$ is an important epidemiological parameter that characterizes the inherent biological transmission of the virus in a completely unprotected population. In particular, it is the mean number of secondarily infected by one infected person introduced in a completely susceptible population. It depends on the biology of the specific virus, as well as the demographic characteristics of the population and the environmental conditions, while it does not depend on the applied infection control measures (Brauer, 2008). In (Salom et al., 2021) we utilized a bioinformatics analysis, akin to those often used to understand complex data in systems biology, to pinpoint demographic and meteorological factors that affect $R_{0,\text{free}}$ (i.e., inherent virus transmissibility in population). This furthermore underlines that a rich array of techniques developed and/or widely used within systems biology can be successfully employed within infectious disease modeling.

### 4. Parameter analysis and inference

$R_{0,\text{free}}$, $\alpha$, $t_0$, two initial conditions ($I_0$ and $E_0$), and the detection efficiency $\varepsilon$, are unknown and may differ between the provinces. Is it possible to determine these unknown parameters from different properties of the $D$ curve? Early in the infection, almost the entire population is susceptible ($S \approx N$), so Eqs. (2) and (3) become linear, and decoupled from the rest of the system, as discussed in the previous section. This sets the ratio of $I_0$ to $E_0$, through the eigenvector components with the dominant (positive) eigenvalue of the Jacobian for this subsystem. This eigenvalue, corresponding to the initial slope of the $log(D)$ curve, sets the value of $\lambda_+$ and subsequently, of $R_{0,\text{free}}$ (see Eq. (11)). From Eq. (7) one can see that the product of $I_0$ and $\varepsilon \cdot \delta$ is set by $dD/dt$ at the initial time ($t=0$). Later dynamics of the $D$ curve is determined solely by the combination $t_\alpha = t_0 + 1/\alpha$ (which we denote as protection time), setting the time at which ~1/2 of the population moves to the protected category. We also numerically checked this, and confirmed that $t_0$ can be lowered at the expense of increasing $1/\alpha$, without affecting the fit quality. We allowed for $t_0$ to vary in reasonable proximity of January 23rd, as the social distancing was generally introduced close to Wuhan closure (e.g., on that date, all major events in Beijing were canceled) (Chen et al., 2020; Du et al., 2020), but we cannot be sure when the measures effectively took place. Our inferred $t_0$ values are within a week from Wuhan closure, appearing as reasonable. The remaining independent parameter ($I_0$) is then left to be determined from $D$ curve properties at the late infection stage, such as its saturation time. The number of characteristic dynamics features is thus at least equal to the number of fit parameters, leading to constrained numerical analysis, so that over-fitting is not expected. For few provinces, we however observed that $I_0$ can be decreased compared to the best fit value, without noticeably affecting the fit quality. For these provinces, we chose the lowest $I_0$ value that still leads to a comparably good fit. This allows obtaining the most conservative (i.e., as high as possible while still consistent with data) IFR estimate, as the reported fatality counts for provinces other than Hubei is surprisingly low.

Parameter inference and uncertainties are estimated separately for each province. However, within a given province, demographic, special, or population activity (network effects) heterogeneities (Britton, Ball, & Trapman, 2020; Diekmann et al., 2012), or seasonality effects (Wong et al., 2020), are not taken into account. These are potentially important, particularly for projections (longer-term predictions of infection dynamics under different scenarios), and can be readily included in our model. Such extensions would however complicate parameter inference, due to an increase in parameter number, as this may either lead to overfitting or require special/additional data that may be available only for a limited number of countries/regions (which would limit the generality of our proposed method). A more complex model structure may also obscure a straightforward relationship between the model parameters and distinct dynamical features of the confirmed case count curve analyzed above. While the inclusion of additional effects is left for future work, we here employ the model structure and parameter inference introduced above on widely available case count data, as proof of the principle for the generality of our proposed approach. Moreover, a major advantage of our approach is that it allows consistent analysis for all provinces with the same model, numerical procedure, and parameter set, allowing an objective comparison of the obtained results.

Our model was numerically solved by the Runge-Kutta method (Dormand & Prince, 1980) for each parameter combination. Parameter values were inferred by exhaustive search over a wide parameter range, to avoid reaching a local minimum of the objective function ($R^2$). To infer the unknown parameters, we fit (by minimizing $R^2$) the model to the observed total number of detected *D* for each province. As an alternative to exhaustive search, some of many optimization techniques used in epidemics modeling, such as the Markov chain Monte Carlo (MCMC) approach, can be used instead (Keeling & Rohani, 2011; Wong et al., 2020) – exhaustive search is however straightforward, guarantees that the global minimum is reached, and is in this case not computationally demanding. Errors were estimated through Monte-Carlo simulations (Press, Flannery, Teukolsky, & Vetterling, 1986), individually for each province with the assumption that count numbers follow the Poisson distribution. Monte-Carlo simulations were found as the most reliable estimate of the fit parameter uncertainties for a non-linear fit (Cunningham, 1993). This also serves as an independent check for overfitting, as in that case, data point perturbations would lead to large parameter uncertainties. We find no indication of this in the results reported below, as the inferred uncertainties (consistently indicated with all results) are reasonably small. In particular, the differences in the inferred parameter values, which are relevant for the reported results/conclusions, are statistically highly significant. *P* values for extracted parameter differences between provinces are estimated by the *t*-test.

## 5. Analysis of COVID-19 transmission in China

We used the SPEIRD with the parameter inference described above, to analyze all Mainland China provinces, except Tibet, where only one COVID-19 case was reported. Parameters were estimated separately for each of the 30 provinces by the same model and parameter set, which enables an impartial comparison of the results presented below. To allow for a straightforward comparison of the infection progression between different provinces, the starting date (i.e., $t=0$) in our analysis is the same for all the provinces and corresponds to January 23$^{rd}$ (when the data for all the provinces became publically available and continuously tracked (Hu et al., 2020)).

In Fig. 4A and B, we show that our model can robustly explain the observed *D*, in the cases of large outburst (Hubei on Fig. 4A), as well as for all other provinces, where *D* is in the range from intermediate (e.g., Guangdong) to low (e.g., Inner Mongolia). Provinces in Fig. 2B were selected to cover the entire range of observed *D* (from lower to higher counts), while comparably good fits were obtained for other provinces, which were all included in the further analysis. Our method is also robust to data perturbations (which might be frequent), e.g. in the case of Hubei (Wuhan), a large number of counts was added on Feb. 12$^{th}$, based on clinical diagnosis (CT scan) (Feng et al., 2020), which is apparent as a discontinuity in observed *D* in Fig. 4A. The model however interpolates this discontinuity, finding a reasonable description of the overall data.

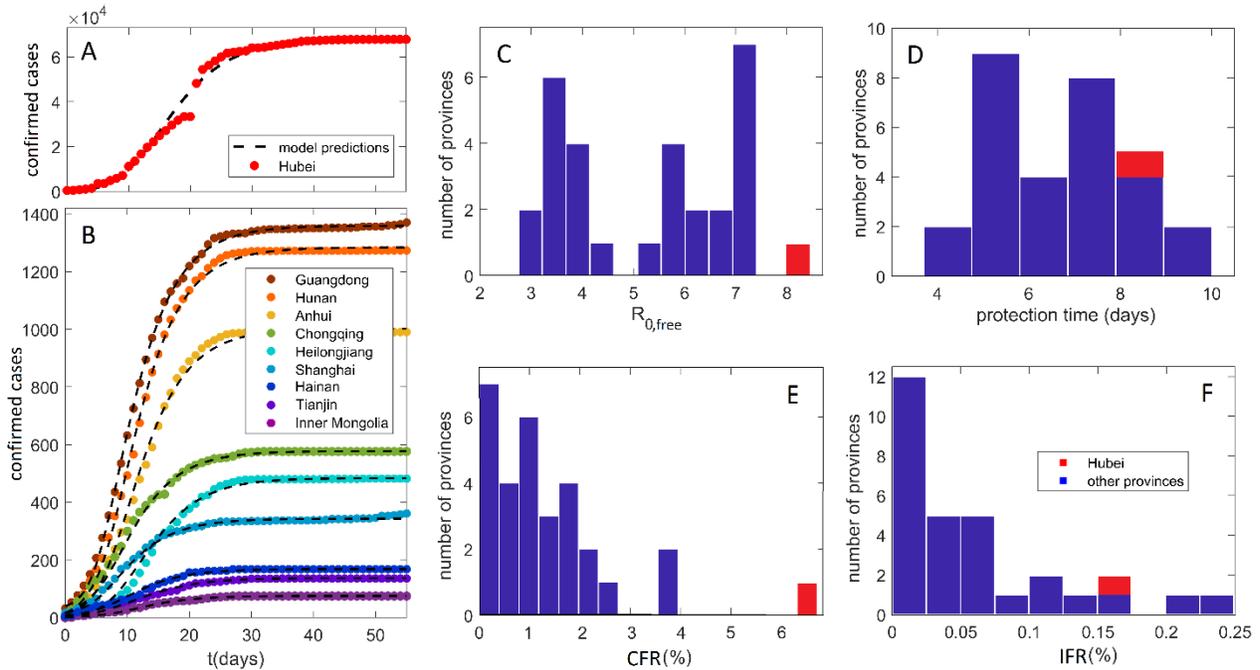

**Fig. 4. Model predictions: comparison with data and key parameter estimates.** Predictions (compared to data) of detected infection counts for (**A**) Hubei, (**B**) other Mainland China provinces. Zeros on the horizontal axis correspond to January 23rd, which is the initial time in our numerical analysis for all the provinces. The observed counts are shown by dots and our model predictions by dashed curves. Names of the provinces are indicated in the legend, with provinces selected to cover the full range of the observed total detected counts. The distribution with respect to provinces of (**C**) the basic reproduction number in the absence of social distancing, $R_{0,free}$, (**D**) the protection time $t_\alpha$. (**E**) Case Fatality Rate, calculated directly from the reported data. (**F**) Infected Fatality Rate. The values for Hubei are indicated by the red bars.

We backpropagated the dynamics inferred for Hubei, to estimate that January 5th (±4 days) was the onset of the infection's exponential growth in the population (not to be confused with the appearance of first infections, which likely happened in December (Feng et al., 2020)). This agrees well with (Feng et al., 2020) (cf. Fig. 3A), which tracked cases according to their symptom onset (shifted for ~12 days with respect to detection/diagnosis, cf. Fig. 3B), and coincides with WHO reports on social media that there is a cluster of pneumonia cases – with no deaths – in Wuhan (WHO, 2020). Since our analysis does not directly use any information before January 23rd, this agreement provides confidence in our $I_0$ estimate. Note that we infer $I_0$ separately for each province of interest, through which we also take into account different times of the infection onset in different provinces (so that earlier onset time would generally lead to a larger number of infected on January 23rd).

| province | $t_\alpha$ (days) | $R_0$ | $E_0$ | $I_0$ | IFR (%) | CFR (%) | detected % |
|---|---|---|---|---|---|---|---|
| Anhui | 6.6±0.5 | 5.5±0.8 | 920±30 | 220±20 | 0.04±0.02 | 0.6±0.3 | 6±3 |
| Beijing | 7.9±0.5 | 3.5±0.4 | 610±20 | 180±10 | 0.12±0.05 | 1.7±0.7 | 7±3 |
| Chongqing | 7.0±0.2 | 3.5±0.2 | 1900±40 | 560±20 | 0.04±0.03 | 1.0±0.5 | 4±2 |
| Fujian | 3.7±0.4 | 7±2 | 1660±40 | 360±20 | 0.007±0.003 | 0.3±0.4 | 2±1 |
| Gansu | 5±1 | 6±3 | 630±20 | 150±10 | 0.03±0.04 | 1±1 | 2±3 |
| Guangdong | 5.0±0.1 | 7±1 | 1360±40 | 290±20 | 0.04±0.01 | 0.6±0.2 | 7±2 |
| Guangxi | 7±1 | 3.8±0.8 | 1000±30 | 290±20 | 0.02±0.02 | 0.8±0.6 | 3±3 |
| Guizhou | 8.1±0.6 | 7±1 | 53±7 | 11±3 | 0.06±0.03 | 1±1 | 4±2 |
| Hainan | 7.6±0.8 | 3.3±0.7 | 300±20 | 90±10 | 0.21±0.09 | 4±2 | 6±3 |
| Hebei | 6.0±0.6 | 7±2 | 240±20 | 52±7 | 0.11±0.03 | 1.8±0.8 | 6±2 |
| Heilongjiang | 7±1 | 6±2 | 260±20 | 59±7 | 0.15±0.07 | 2.9±0.9 | 5±3 |
| Henan | 7.0±0.3 | 4.5±0.5 | 1780±40 | 460±20 | 0.09±0.04 | 1.7±0.4 | 5±2 |
| Hubei | 8.3±0.2 | 8.2±0.4 | 31900±400 | 6600±200 | 0.15±0.09 | 6.5±0.1 | 2±2 |
| Hunan | 5.1±0.1 | 6.8±0.8 | 1430±40 | 310±20 | 0.02±0.01 | 0.4±0.2 | 5±2 |
| I. Mongolia | 10.0±0.8 | 2.8±0.4 | 940±30 | 300±20 | 0.01±0.03 | 1±1 | 1±3 |
| Jiangsu | 5.5±0.5 | 7±2 | 500±20 | 110±10 | 0±0 | 0±0 | 6±2 |
| Jiangxi | 7.0±0.2 | 5.6±0.9 | 890±30 | 210±10 | 0.005±0.002 | 0.1±0.1 | 5±2 |
| Jilin | 10.0±0.7 | 4.0±0.8 | 270±20 | 76±9 | 0.02±0.02 | 1±1 | 1±2 |
| Liaoning | 7±1 | 2.9±0.7 | 1240±40 | 390±20 | 0.02±0.04 | 2±2 | 1±2 |
| Ningxia | 5.3±0.9 | 7±3 | 72±9 | 15±4 | 0±0 | 0±0 | 6±23 |
| Qinghai | 6.1±0.6 | 4.0±0.5 | 2260±50 | 640±30 | 0±0 | 0±0 | 0±2 |
| Shaanxi | 5.2±0.5 | 6±1 | 380±20 | 90±10 | 0.07±0.03 | 1.3±0.8 | 6±2 |
| Shandong | 9±1 | 3.5±0.5 | 900±30 | 260±20 | 0.06±0.01 | 1.0±0.4 | 6±1 |
| Shanghai | 5.0±0.4 | 6±1 | 1570±40 | 370±20 | 0.02±0.02 | 0.8±0.5 | 2±3 |
| Shanxi | 5.2±0.5 | 6±2 | 1600±40 | 370±20 | 0±0 | 0±0 | 1±2 |
| Sichuan | 7.7±0.8 | 3.7±0.5 | 990±30 | 280±20 | 0.03±0.02 | 0.6±0.3 | 5±3 |
| Tianjin | 7±2 | 4±2 | 170±10 | 46±7 | 0.14±0.06 | 2±1 | 7±3 |
| Xinjiang | 7.3±0.9 | 6±1 | 42±7 | 10±3 | 0.25±0.09 | 3±2 | 8±2 |
| Yunnan | 4.0±0.2 | 7±2 | 360±20 | 76±9 | 0.06±0.03 | 1.2±0.9 | 5±2 |
| Zhejiang | 5.0±0.1 | 7.2±0.8 | 1340±40 | 290±20 | 0.005±0.002 | 0.1±0.1 | 7±3 |

**Table 1:** $t_\alpha$ – protection time, $R_{0,\text{free}}$ – basic reproduction number, $E_0$ – initial exposed, $I_0$ – initial infected, IFR – Infection Fatality Rate, CFR – Case Fatality Rate, detected % - fraction of the infected population that has been detected. Error of the quantities correspond to one standard deviation.

Key parameters inferred from our analysis are summarized in Fig. 4C-F, with individual results and errors for all the provinces shown in the Table 1. Fig. 4C shows the distribution of $R_{0,\text{free}}$. Note that $R_{0,\text{free}}$ might depend on demographic (population density, etc.) and climate factors (temperature, humidity…), which are not controllable, but are unrelated to the applied social distancing measures (see above). It is known that the $R_0$ value can strongly depend on the model, e.g. the number of introduced compartments (Keeling & Rohani, 2011); accordingly, a wide range of $R_0$ values were reported for China in the literature (Sanche et al., 2020; Wu, Leung, Bushman, et al., 2020). Consequently, a clear advantage of our study is that parameters for all China provinces were determined from the same model and data set, which allows direct comparisons. Our obtained average $R_{0,free}$ for provinces outside of Hubei is 5.3±0.3, in a reasonable agreement with a recent estimate (≈5.7) (Sanche et al., 2020). Furthermore, we observe that $R_{0,free}$ for Hubei is a far outlier with a value of 8.2±0.4, which is notably larger than for other provinces with p~$10^{-11}$. This then strongly suggests that demographic and climate factors that determine $R_{0,free}$, played a decisive role in a large outburst in Hubei vs. other provinces, which we further address below.

The distribution of protection time $t_\alpha$ for the provinces is shown in Fig. 4D, with the value for Hubei indicated in red. The mean for the other provinces is 6.6±0.2 days. That is, we observe that the suppression measures were efficiently implemented, with ~½ of the population moving to the protected category within a week from Wuhan closure. The protection time for Hubei of 8.3±0.2 days was longer, which is statistically significant at the p~$10^{-11}$ level. The estimated less efficient protection in the case of Hubei may also be an important contributing factor in the surprising difference in Hubei vs. other provinces, which we further investigate below.

CFR distribution, based on the fatality numbers reported for Hubei and other provinces is shown in Fig. 4E. These numbers are not based on the model predictions, i.e., can be straightforwardly obtained by dividing the total number of fatalities by the total number of detected cases. CFR for other provinces with a mean of 1.2±0.4% is significantly smaller compared to CFR for Hubei, which was 4.6% before the correction on April 17th, and 6.5% after the correction (with 1290 fatalities added to Wuhan). This large difference in CFR between Hubei and other provinces further accentuates the differences noted in Fig. 1.

IFR is harder to determine than CFR, as a majority of COVID-19 infections correspond to asymptomatic or mild cases that are by large not diagnosed (Day, 2020). We consequently calculate IFR as the total number of fatalities divided by the total number of infections (cumulative incidence) for the entire outburst, where cumulative incidence is estimated from our model. As the infections precede fatalities, both the total number of fatalities and the cumulative incidence in our estimate correspond to the entire outburst, so that all the infections had a sufficient time to recover or lead to fatalities – this is directly feasible for the provinces in China, where all detected case counts reached saturation. Note that IFR calculated in this way corresponds to an averaged quantity so that it does not capture possible time-dependent change over the outburst interval (in fact, for Wuhan it is known that the fatality rate was larger at the very beginning of the outburst). Nevertheless, the estimated IFR's present a reasonable measure of COVID-19 mortality across China provinces.

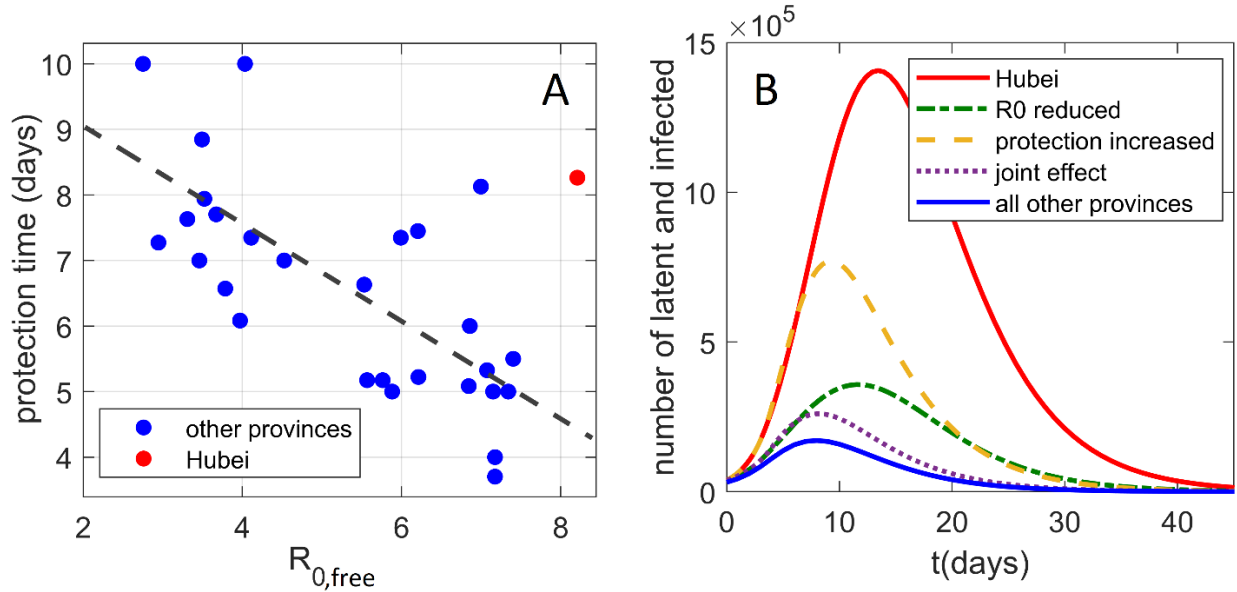

**Fig. 5. The interplay of transmissibility and effective social distancing.** (**A**) The correlation plot of $t_\alpha$ vs. $R_{0,\text{free}}$ for all provinces, where the point corresponding to Hubei is marked in red. (**B**) The effect (on the Hubei dynamics of infected and latent cases) of reducing $R_{0,\text{free}}$ and $t_\alpha$ to the mean values of other Mainland China provinces. Both the unperturbed Hubei dynamics and the sum of infected and latent cases for all other provinces are included as references.

IFR distribution, which provides a much less biased measure of the infection mortality, is shown in Fig. 4F. In distinction to CFR, estimated IFR shows a much smaller difference between Hubei (0.15%±0.09%) and other provinces (0.056±0.007%). Therefore, while Hubei is a clear outlier with respect to CFR, we observe similar IFR values for all Mainland China provinces, where few provinces have even higher IFR than Hubei. The ratio of IFR to CFR equals the fraction of all infected that got detected (*detection coverage*). We estimate that the mean detection coverage for all provinces except Hubei is higher than detection coverage for Hubei (4.5±0.9% vs. 2±2%). This difference is responsible for a decrease by a factor of two from CFR to IFR for Hubei, compared to the other provinces, and consequently for more uniform mortality estimates at the IFR level. Xinjiang has the highest IFR of 0.25±0.09% so that Hubei is not an outlier anymore. Estimated IFR's of up to 0.3% in China provinces are in general agreement with the estimates reported elsewhere (see e.g. (Bar-On et al., 2020; Magdalena Djordjevic et al., 2021; Mizumoto, Kagaya, & Chowell, 2020).

In Fig. 5A, two key infection progression parameters are plotted against each other: protection time $t_\alpha$ vs. basic reproduction number $R_{0,\text{free}}$. Unexpectedly, there is a high negative correlation, with Pearson correlation coefficient R = − 0.70, which is statistically highly significant p~$10^{-5}$, where these two are *a priori* unrelated (see above). Actually, stronger social distancing measures - which by definition are not included in $R_{0,\text{free}}$ - would lead to a decrease in *effective* transmissibility. This would then lead to a tendency of transmissibility to positively correlate with $t_\alpha$, oppositely from the strong negative correlation observed in Fig. 5A. Therefore, higher basic reproduction number is genuinely related to a shorter protection time (larger effect of the suppression measures). Intuitively, this could be understood as a negative feedback loop, commonly observed in systems biology (Alon, 2019; Phillips, Kondev, Theriot, & Garcia, 2012), where larger $R_{0,\text{free}}$ leads to steeper initial growth in the infected numbers, which may elicit stronger measures and better observing of these measures by the population faced with a more serious outbreak. Interestingly, similar negative feedback was also obtained in the context in the context of epidemics research other than COVID-19 (Wang, Andrews, Wu, Wang, & Bauch, 2015).

The two main properties of the Hubei outburst are therefore higher $R_{0,\text{free}}$ and $t_\alpha$ compared to other provinces. In Fig. 5B, we investigate how these two properties separately affect the Wuhan outburst for latent and infected cases, where unperturbed Hubei dynamics is shown by the red full curve. We first reduce only $R_{0,\text{free}}$ from the Hubei value, to the mean value for all other provinces (the dash-dotted green curve). We see that this reduction substantially lowers the peak of the curve, though it still remains wide. Next, instead of decreasing $R_{0,\text{free}}$, we decrease the protection time $t_\alpha$ to the mean value for all other provinces (dashed orange curve). While reducing $t_\alpha$ also significantly lowers the peak of the curve, its main effect is in narrowing the curve, i.e., reducing the outburst time. Finally, when $R_{0,\text{free}}$ and $t_\alpha$ are jointly reduced, we obtain the (dotted purple) curve that is both significantly lower and narrower than the original Hubei progression. This curve comes quite close to the curve that presents the sum of all other provinces (full blue curve) - the dotted curve remains somewhat above this sum, mainly because the initial number of latent and infected cases is somewhat higher for Hubei compared to the sum of all other provinces. This synergy between the transmissibility and the control measures will be further discussed below.

## 6. Conclusions

In this study, we applied a systems biology approach to develop a novel method of COVID-19 transmission dynamics. The model includes (time-dependent) social distancing measures in a simple manner, consistent with the compartmental mechanistic nature of the underlying process. The model has a major advantage that it is independent of the specific transmission process considered, and requires only commonly available count data as an input. The model allows extracting key infection parameters from the data that are readily available and publicly accessible (both for China and other countries), so that, in a nutshell, our approach is of wide applicability. To our best knowledge, such parameters (necessary to assess any future COVID-19 risks), were not extracted by other computational approaches.

The developed method is subsequently applied to the problem that appears highly non-trivial, i.e., to understand the puzzle created by the drastic differences in the infection and fatality counts between Hubei and the rest of Mainland China. The goal was to determine if it is possible to consistently explain such drastic differences by the same model, and what are the resulting numerical estimates and conclusions. We found that Hubei was a suitable ground for infection transmission, being an outlier with respect to two key infection progression parameters: having significantly larger $R_{0,\text{free}}$, and a longer time needed to move a sizable fraction of the population from susceptible to a protected category. While stricter measures were formally introduced in Hubei, the initial phase of the outburst put a large strain on the system, arguably leading to less effective measures compared to other provinces.

The fact that the initial epidemic in Hubei was not followed by similar outbursts in the rest of Mainland China may be understood as a serendipitous interplay of the two factors noted above. While both smaller $R_{0,\text{free}}$ and lower half-protection time (more efficient measures) significantly suppress the infection curve, their effect is also qualitatively different. While lowering $R_{0,\text{free}}$ more significantly suppresses the peak, decreasing the half-protection time significantly reduces the outburst duration. Consequently, the synergy of these two effects appears to lead to drastically suppressed infection dynamics in other Mainland China provinces compared to Hubei. The number of detected (diagnosed) cases in the entire Mainland China is, therefore, though unintuitive, well consistent with the model, and is explainable by a seemingly reasonable combination of circumstances. Our obtained negative feedback between transmissibility and effects of social distancing may be understood in terms of larger transmissibility triggering more stringent social distancing measures, where a similar conclusion was also obtained through entirely different means (a combination of real-time human mobility data and regression analysis) (Kraemer et al., 2020).

In summary, we showed that unintuitive dissimilarity in the infection progression for Hubei vs. other Mainland China provinces is consistent with our model, and can be attributed to the interplay of transmissibility and effective protection, demonstrating that regional differences may drastically shape the infection outbursts. This also shows that comparisons in terms of the confirmed cases, or fatality counts (even when normalized for population size), between COVID-19 and other infectious diseases, or between different regions for COVID-19, are not feasible, and that parameter inference from quantitative models (individually for different affected regions) is necessary. Consequently, this paper illustrates that utilization of uncommon strategies, such as systems biology application to mathematical epidemiology, may significantly advance our understanding of COVID-19 and other infectious diseases.

**Acknowledgements:** This work was supported by the Ministry of Education, Science and Technological Development of the Republic of Serbia.